# A modular phantom and software to characterize 3D geometric distortion in MRI

Jordan M. Slagowski[1], Yao Ding[2], Manik Aima[1], Zhifei Wen[1], Clifton D. Fuller[2], Caroline Chung[2], J. Matthew Debnam[3], Ken-Pin Hwang[4], Mo Kadbi[5], Janio Szklaruk[6], and Jihong Wang[1]

[1]Department of Radiation Physics, The University of Texas MD Anderson Cancer Center, Houston, TX 77030, USA

[2]Department of Radiation Oncology, The University of Texas MD Anderson Cancer Center, Houston, TX 77030, USA

[3]Department of Neuroradiology, The University of Texas MD Anderson Cancer Center, Houston, TX 77030, USA

[4]Department of Imaging Physics, The University of Texas MD Anderson Cancer Center, Houston, TX 77030, USA

[5]MR Therapy, Philips HealthTech, Cleveland, OH, United States

[6]Department of Abdominal Imaging, The University of Texas MD Anderson Cancer Center, Houston, TX 77030, USA

**Running Title**

Phantom to Characterize Distortion in MRI

**Correspondence to:**

Jihong Wang, Ph.D.

Department of Radiation Physics, The University of Texas MD Anderson Cancer Center

1515 Holcombe Blvd, Unit 94, Houston, TX 77030, USA.

Tel: (713) 563-2531; E-mail: Jihong.Wang@mdanderson.org

**Abstract**

Magnetic resonance imaging (MRI) offers outstanding soft tissue contrast that may reduce uncertainties in target and organ-at-risk delineation and enable online adaptive image-guided treatment. Spatial distortions resulting from non-linearities in the gradient fields and non-uniformity in the main magnetic field must be accounted for across the imaging field-of-view to prevent systematic errors during treatment delivery. This work presents a modular phantom and software application to characterize geometric distortion (GD) within the large field-of-view MRI images required for radiation therapy simulation. The modular phantom is assembled from a series of rectangular foam blocks containing high-contrast fiducial markers in a known configuration. The modular phantom design facilitates transportation of the phantom between different MR scanners and MR-guided linear accelerators and allows the phantom to be adapted to fit different sized bores or coils. The phantom was evaluated using a 1.5T MR-guided linear accelerator (MR-Linac) and 1.5T and 3.0T diagnostic scanners. Performance was assessed by varying acquisition parameters to induce image distortions in a known manner. Imaging was performed using T1 and T2 weighted pulse sequences with 2D and 3D distortion correction algorithms and the receiver bandwidth (BW) varied as 250-815 Hz/pixel. Phantom set-up reproducibility was evaluated across independent set-ups. The software was validated by comparison with a non-modular phantom. Average geometric distortion was $0.94 \pm 0.58$ mm for the MR-Linac, $0.90 \pm 0.53$ mm for the 1.5 T scanner, and $1.15 \pm 0.62$ mm for the 3.0T scanner, for a 400 mm diameter volume-of-interest. GD increased, as expected, with decreasing BW, and with the 2D versus 3D correction algorithm. Differences in GD attributed to phantom set-up were 0.13 mm or less. Differences in GD for the two software applications were less than 0.07 mm. A novel modular phantom was developed to evaluate distortions in MR images for radiation therapy applications.

## 1. Introduction

Magnetic resonance imaging (MRI) offers superior soft tissue contrast comparatively to x-ray based computed tomography (CT), making it an attractive imaging modality for patient simulation and image guided radiation therapy (Verellen *et al* 2007). The registration or fusion of an MR image with a CT image has been shown to aid in tumor and organ-at-risk (OAR) contouring (Verellen *et al* 2007, Debois *et al* 1999). In certain scenarios, synthetic CT images can be generated directly from MR images, thereby replacing the need for a simulation CT altogether and reducing the associated uncertainty attributed to image registration (Kim *et al* 2015, Hsu *et al* 2013, Johnstone *et al* 2017), making an MRI-only workflow for treatment planning possible. The recent integration of an MRI scanner with a linear accelerator (Raaymakers *et al* 2009, Mutic and Dempsey 2014) offers improved online visualization of low-contrast lesions in the abdominal or head and neck regions relative to cone beam CT, as well as real-time tumor tracking (Yun *et al* 2013), and a daily adaptive treatment capability (Pathmanathan *et al* 2018). Each of these applications of MRI to radiotherapy requires that the 3D image volumes be acquired with high spatial fidelity and geometric accuracy. However, geometric distortion of MR images is a well-known shortcoming that may occur due to patient-specific or MRI system hardware-specific factors (Weygand *et al* 2016). Patient-specific sources of geometric distortion include chemical shift and magnetic susceptibility (Stanescu *et al* 2012). Hardware-specific factors can vary from scanner to scanner and include non-uniformity in the main magnetic field or non-linearities in the gradient fields used for spatial encoding. The dosimetric impact of image distortions on plan quality was quantified in a recent study that showed deviations in maximum dose of up to 5.2 Gy and a 3.7% loss of target coverage (Yan *et al* 2018). A thorough review of the sources of geometric distortion and techniques to measure and correct for distortion is provided by Weygand *et al.* (2016).

The objective of phantom based distortion measurement techniques is to determine the spatial deviation of landmark objects or features identified in an MR image from known reference positions. Task Group 1 of the American Association of Physicists in Medicine (AAPM TG-1) recommended that distortion phantoms occupy a large field-of-view and consist of regularly spaced objects including holes, grooves, rods or tubes (Price *et al* 1990). AAPM TG-1 suggested two example phantom designs including orthogonal grooves in an acrylic plate or an array of holes drilled in an acrylic plate filled with an MR-visible material. Early designs included a 2D phantom consisting of a water-filled cylinder (34 cm diameter, 10 cm height) with a rectangular grid of plastic rods and a 3D plastic cylinder (26 cm diameter, 8 cm height) with water-filled holes (Price *et al* 1990, Schad *et al* 1987). Since then, a number of other phantom designs have been reported including an acrylic grid-pattern phantom (24 cm x 24 cm x 20 cm) (Schad *et al* 1992), a 3D printed grid phantom (17 cm x 17 x cm x 25 cm) (Mizowaki *et al* 2000), and an acrylic box (15 cm x 15 cm x 18 cm) filled with acrylic rods (Jafar *et al* 2017, Yu *et al* 2001). Limitations of applying these early phantom designs for MRI in RT quality assurance include the small dimensions, fixed maximum dimension, and heavy weight. In contrast to diagnostic MR imaging, images acquired for radiotherapy require geometric distortion to be assessed for the variable, and larger field-of-view (FOV) required as input by treatment planning systems for dose calculations.

A larger cube phantom measuring 31 cm by 31 cm by 31 cm and consisting of a series of plastic grid sheets filled with a water solution was fabricated by *Wang and Doddrell* (2004). Despite being larger than previous phantoms, the phantom by *Wang and Doddrell* would still not cover the full FOV of certain scanners, such as the 50 cm by 50 cm by 45 cm FOV of the Siemen's MAGNETOM Aera. Increasing the size of the previously proposed phantoms may be an option, however, one must also consider phantom portability and weight in phantom design. Larger phantoms are also more susceptible to damage and may be more difficult to repair. Furthermore, it is desirable that a single geometric distortion phantom could be used for the different types of MR scanners and bore geometries at a given institution. A modular phantom could be used to measure distortion within the full MRI field-of-view while remaining portable, compatible with different bore sizes, and easier to repair. Price *et al.* (2017) recently reported a modular phantom design consisting of a series of polyurethane foam *plates* with six mm paintball inserts.

The purpose of this work was to develop and validate a novel modular-based *block* phantom and software application to characterize geometric distortion in MRI for radiation therapy simulation or image guidance. The developed phantom, described in the following section, consists of a series of small (10 cm by 10 cm) lightweight blocks that are assembled to cover a larger FOV. The major advantage of the proposed *block* design versus the *plate* design is the increased degrees-of-freedom for phantom configuration. For example, the block phantom can be adapted to evaluate geometric distortion for different sized radiofrequency (RF) receiver coils placed near the target volume. The plate phantom was evaluated for a body coil but would not be compatible with smaller head or extremity RF coils due to the larger width and length of the plates. The current body coil available for the MR-linac has several limitations that motivate the development of new RF coils (Hoogcarspel *et al* 2018). New coils will require new tools such as the modular block phantom to assess the impact on spatial distortions. A second advantage of using blocks is that a small block can be more easily repaired or replaced than the larger plates or phantoms. A third novel feature of the described phantom is that the MR visible markers are contained in exchangeable plugs that are inserted in each block. Different types of exchangeable plugs are under development to evaluate image quality (e.g. uniformity and resolution) at different locations within the MR bore. Finally, the block design allows for Radiochromic film to be placed vertically or horizontally between bricks which may be used to evaluate radiation and imaging coincidence as a function of location in the bore and/or RF coil.

## 2.  Methods and Materials

### 2.1 Phantom design

A modular, large field-of-view geometric distortion phantom (Figure 1) containing 828 high-contrast fiducials made of paintballs, suspended in a foam material was fabricated in-house. The foam material is not visible in MRI but the fiducials are. The diameter of each spherical fiducial measures 6 mm. The fiducials are spaced 50 mm in each of the x-, y-, and z- dimensions corresponding to the medial-lateral (ML), anterior-posterior (AP), and superior-inferior (SI) axes, respectively. The coordinate system is defined with respect to the MR bore as shown in Figure 1. The phantom was designed to span a large volume-of-interest (50 cm lateral, 40 cm anterior-posterior, 40 cm cranial-caudal) in order to assess distortion that may occur near the skin surface or at regions distant from the center of the MR bore. As shown in Figure 1, the phantom is composed of a series of 18 individual foam blocks in order to facilitate transportation between different scanners. The modular block design allows the phantom to be adapted to fit MR bores of different dimensions. Additional blocks could be added if needed to assess distortion within a larger bore or FOV. The dimensions of an individual block measure 10 cm in the x-dimension (width), 10 cm in the y-dimension (height), and 40 cm in the z-dimension (length). Each of the foam blocks contains a total of 46 fiducials. The fiducials are contained within exchangeable plugs that are inserted in the foam blocks. A combination of markings and labels on each of the blocks are used during phantom set-up to improve reproducibility.

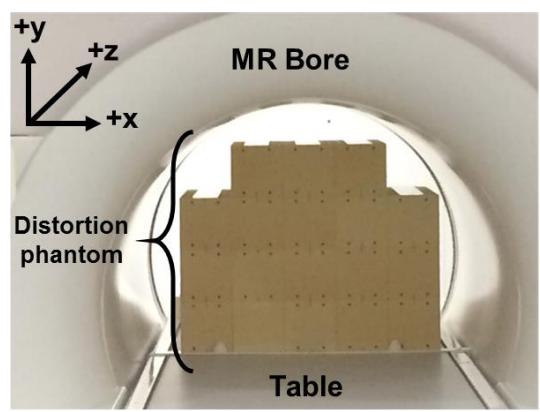

Figure 1: The geometric distortion phantom and orientation of the reference coordinate system are shown within the bore of an MRI scanner.

## 2.2 Reference phantom model

To quantify geometric distortion in a volume of MRI images the position of each high-contrast fiducial in the images must be compared versus the actual position of the fiducial in the phantom (i.e the reference position). A digital phantom, also referred to as a reference model, of the physical phantom was generated based on the manufacturing plans used to fabricate the phantom. The reference model specifies the three-dimensional coordinates of each fiducial position within the physical phantom. The phantom manufacturing tolerances contain an uncertainty of ±0.1 mm in fiducial position. A CT scan of the phantom was performed (SOMATOM Definition Edge, Siemens Healthcare, Forchheim, Germany) at 120 kVp and 500 mAs in order to validate the reference model matched the physical phantom. The reconstruction field-of-view was set at 65 cm in order to minimize truncation artifacts. The reconstructed image voxel dimensions were 1.27 mm by 1.27 mm in the x- and y-dimensions, respectively. The slice thickness was 1.00 mm in the z-dimension. Example axial and coronal CT images of the phantom are presented in Figure 2.

Spatial coordinates for each of the fiducials were determined through a two-step procedure. First, fiducial coordinates were initialized as the expected coordinates according to the reference model. Second, an overlay of each of the fiducial coordinates on the CT image volume was reviewed and the coordinates were manually adjusted to correspond to the center of the fiducial. The mean distance and standard deviation (σ) between the reference model and CT-derived coordinates were 0.19 mm and 0.53 mm, respectively. Uncertainty in the CT-derived coordinates was comparable to the 0.1 mm manufacturing tolerance. As the markers were designed to be visible under MR imaging it was difficult to automatically determine the position of the fiducials within the CT reference volume using an intensity-based thresholding method. The reference model selected and used throughout this work was that derived from the manufacturing specifications due to a lower uncertainty (0.1 mm) versus the CT-derived model (0.19 mm). Future work may consider the design and use of a fiducial material that is visible under both CT and MR imaging to reduce uncertainty in the CT derived reference model.

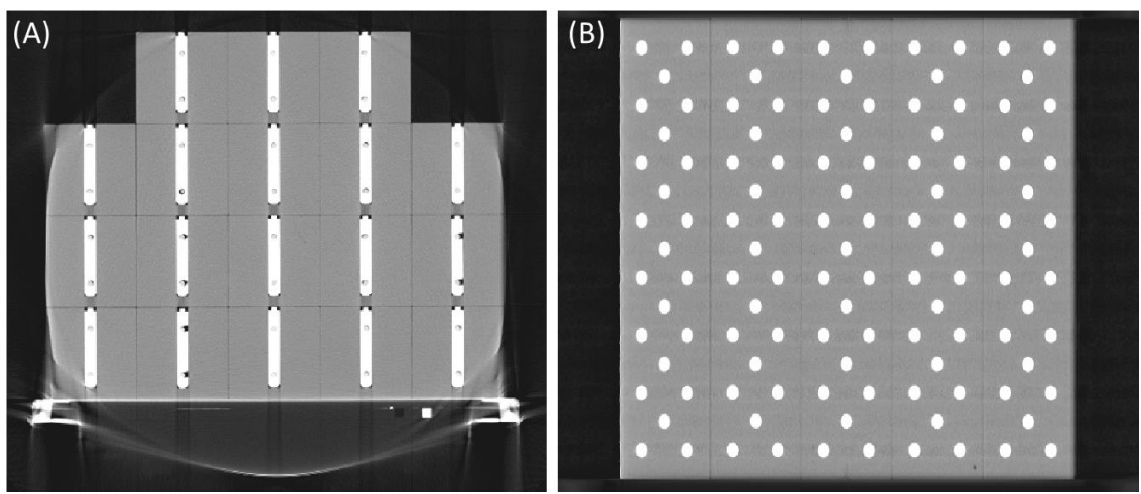

Figure 2: Reference computed tomography images for axial (left) and coronal slices (right) are shown for the geometric distortion phantom.

## 2.3 Geometric distortion assessment

Geometric distortion was assessed using software developed in-house with the MATLAB programing language (MathWorks, Inc., Natick, Massachusetts) based on the methodology described in this section.

### 2.3.1 Definition of geometric distortion

Three-dimensional geometric distortion within the MR bore was characterized in a manner similar to that described by *Wang and Doddrell* (2004). The spatial coordinates defining the position of a reference fiducial, *i,* within the reference coordinate system (Figure 1), were specified as $(x_i^{ref}, y_i^{ref}, z_i^{ref})$. The complete collection of reference coordinates (N = 824) were stored in an N x 3 matrix, $\mathbf{q^{ref}}$, where the $i^{th}$ element $q_i^{ref}$ describes the position of fiducial *i.* After MR imaging of the geometric distortion phantom, a series of spatial coordinates describing the measured location of each fiducial *i* was determined and represented as $(x_i, y_i, z_i)$. The complete collection of measured fiducial coordinates were stored in a matrix $\mathbf{q.}$ The measured fiducial coordinates $\mathbf{q}$ were then registered with the reference coordinates $\mathbf{q^{ref}}$ using a rigid registration technique (Besl *et al* 1992). The rigid registration accounts for small phantom rotations or translations that may occur during phantom set-up. The registration method was constrained to use only fiducials within an 11.0 cm radius in the XY plane and ± 3.0 cm about magnetic isocenter in the z-dimension. It is expected that geometric distortion is minimal at such short distances from the magnetic isocenter (Torfeh *et al* 2016). Geometric distortion was characterized by a set of distortion vectors, $\mathbf{d}$, describing the direction and magnitude of displacement of a measured fiducial position from the expected fiducial position. The components of vector $\mathbf{d}$ for fiducial *i* are defined as,

$$d_{x,i} = x_i - x_i^{ref},$$

$$d_{y,i} = y_i - y_i^{ref}, \hspace{4cm} (1)$$

$$d_{z,i} = z_i - z_i^{ref},$$

and describe the direction and magnitude of geometric distortion in the ML, AP, and SI dimensions, respectively. The absolute geometric distortion, $d_i$, was computed as the Euclidean distance between a measured and reference fiducial position according to equation 2.

$$d_i = \sqrt{d_{x,i}^2 + d_{y,i}^2 + d_{z,i}^2}$$

(2)

### 2.3.2 Geometric distortion characterization

Geometric distortion is reported in terms of the mean and standard deviation of the measured distortion in each of the ML, AP, and SI dimensions. The absolute distortion is also reported. Statistics are provided for fiducials contained within the full imaging field-of-view, as well as regions-of-interest. Specifically, results are reported for fiducials within spherical regions-of-interest centered at isocenter with diameters of 200 mm, 300 mm, 400 mm, and 500 mm. The maximum distortion value and distortion value corresponding to the 98[th] percentile of a histogram of the distortion measurements are also reported. A series of 2D and 3D heat maps provide visualization of the distortion vectors as a function of spatial location within the MR bore.

### 2.3.3 Fiducial detection algorithm

Each of the fiducial coordinates $(x_i, y_i, z_i)$ acquired from an MR image volume of the distortion phantom were determined using a semi-automatic approach. First, image processing and segmentation were performed to determine the coordinates of each fiducial. The segmented fiducials were then fused with the acquired MR image volume and individual image slices were reviewed to confirm that each of the fiducials was properly identified. Manual intervention was performed as needed to correct for any false positives or false negatives. The complete segmentation procedure is described in detail here.

Figure 3A presents an example T1-weighted image of the geometric distortion phantom. The image display window was set between the minimum voxel value and 15% of the maximum voxel value in order to emphasize contrast differences across the FOV. Due to contrast variation across the FOV, global intensity-based thresholding failed to provide accurate fiducial segmentation. Image processing was performed to improve the uniformity across the imaging FOV, reduce image noise, and enhance fiducial contrast in order to facilitate segmentation. In the proposed scheme, each image slice was first convolved with a two-dimensional boxcar filter measuring 20 mm by 20 mm. The filtered image was subtracted from the original image in order to suppress the image background and enhance local contrast, as demonstrated in Figure 3B. Next, the contrast-enhanced image volume was convolved with a 4 mm diameter pillbox (i.e. disk) kernel to identify the circular fiducial objects in the scene. Figure 3C presents an example image after the pillbox filter was applied. At this stage, an intensity-based thresholding method could be deployed to segment the fiducial objects. However, improved robustness in the segmentation process across different pulse sequences and acquisition parameters was observed after performing three additional filter operations prior to segmentation. The additional image processing included filtering the image with horizontal and vertical Prewitt kernels to detect the edges of the fiducial objects. A second, 6 mm diameter, pillbox kernel spreads the high gradient edges across the fiducial object area. Finally, after performing each of the filtering steps, a binary image identifying each of the fiducials was obtained by performing intensity-based thresholding. Voxels with values less than 10% of the maximum image value were set to zero while voxels greater than or equal to 10% of the maximum value were set to one. Following segmentation, a connected components analysis was performed to label each of the detected fiducials. The volume of each detected object was computed and objects with a volume less than 80% of the expected fiducial volume were removed. Filtering the objects by volume was found to reduce the number of false positives that may otherwise appear near the periphery of the image volume. The 3D spatial coordinates of each fiducial were initialized as the centroid of the binary fiducial objects. Figure 3D presents an example of the detected fiducial locations, identified by yellow circles, overlaid on the filtered image.

Each image and segmentation result were manually inspected to ensure the fiducials in low-contrast regions, often near the edges of the imaging volume where distortion is greatest, were detected. An example image demonstrating a fiducial that was not correctly identified is shown in Figure 4. A MATLAB based graphical user interface (GUI) was developed to review each of the MR image slices and superimposed 3D fiducial coordinates. The developed GUI provides a pair of yellow cross-hairs that were positioned by the user over any fiducials not properly identified by the segmentation algorithm. The user can add missing fiducials to the vector **q** with a left-click of the mouse. Similarly, incorrectly identified fiducials may be removed with a right-click of the mouse. After the manual review and editing procedure, a refined set of fiducial coordinates $(x_i, y_i, z_i)$ was computed for each fiducial as the weighted grayscale center-of-mass (C.O.M.) within a 10 mm by 10 mm by 8 mm volume-of-interest centered about the initial fiducial C.O.M. determined from the binary image.

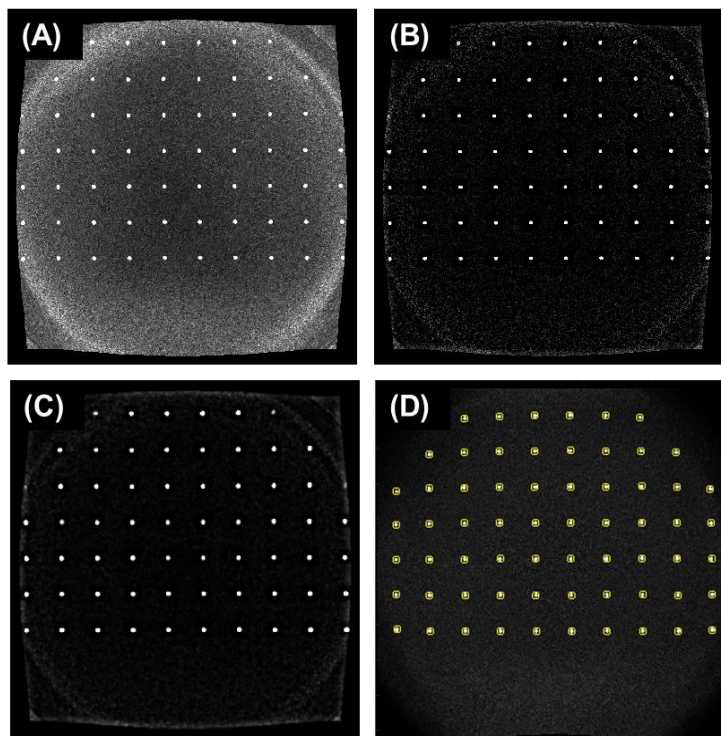

Figure 3: An example T1-weighted MR image of the modular geometric distortion phantom is shown (A). Convolution of the original image with a 2D boxcar kernel improves image contrast (B). The combination of a disk kernel and Prewitt kernel are used to identify fiducials in the image (C). The detected fiducial coordinates are overlaid on the MR image for visualization (D).

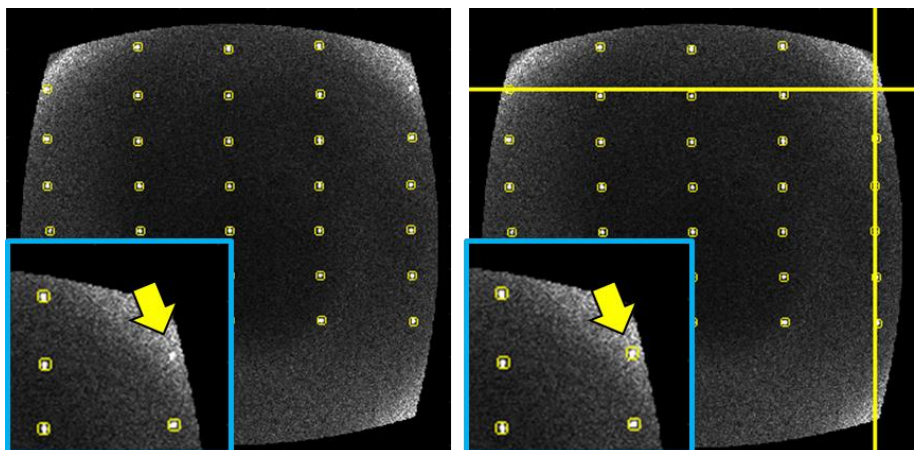

Figure 4: An example MR image of the geometric distortion phantom is shown on the left. The magnified region-of-interest and yellow arrow demonstrate a fiducial that was not properly identified by the automatic segmentation algorithm. A GUI (right image) allows the user to select or delete fiducials that were not properly labeled.

## 2.4 Evaluation of software accuracy

In order to validate the accuracy of the developed software application, geometric distortion was assessed for an MR scan of a second geometric distortion phantom and the analysis performed using our developed software was compared versus results from an independent software application. The 3D Geometric QA Phantom (3D Geometric QA Phantom, Philips Medical Systems MR, Vantaa, Finland) developed for non-clinical research by Philips Medical Systems was scanned at our institution on a 1.5T MR-Linac. The phantom contains a series of oil capsules distributed over a cylindrical volume with a diameter of 500 mm and a length 330 mm. The markers are spaced 25 mm by 25 mm by 55 mm. The phantom was scanned with a T1 weighted (TR = 6.7 ms, TE = 3.4 ms) pulse sequence and bandwidth of 431 Hz/pixel. The diameter of the reconstructed field-of-view measured 560 mm with 1.09 mm by 1.09 mm voxel dimensions and 2 mm slice thickness. Geometric distortion was analyzed for the acquired MR images using the vendor provided 3D Geometric QA Analysis tool. The vendor provided software quantifies geometric distortion for each of the markers and statistics for volumes-of-interest defined about isocenter. Geometric distortion was determined for fiducials contained within spherical volumes centered about iso-center with diameters of 200 mm, 300 mm, 400 mm, and 500 mm. Absolute distortion was specified for each of the volumes-of-interest in terms of the maximum distortion distance and also the distance corresponding to the 98th percentile of a histogram of individual marker distortion distances. The maximum distortion and 98th percentile distortion were computed for the medial-lateral, anterior-posterior, and superior-inferior dimensions. The software developed as part of this work was applied to evaluate geometric distortion for the same image series and compared versus the vendor's results.

## 2.5 Reproducibility of phantom set-up

As discussed, the phantom consists of a series of 18 blocks that are arranged according to a sequence of markings and labels to improve the reproducibility of phantom set-up. To quantify the uncertainty that can be attributed to variations in phantom set-up, the distortion phantom was imaged three times using a T1-weighted pulse sequence. Each of the imaging studies was performed using the same pulse sequence and scanner. The phantom was removed from the scanner table and reassembled before each scan making the set-up of each acquisition independent of the other acquisitions. Geometric distortion was computed and compared across acquisitions to determine the component of uncertainty that can be attributed to set-up variation. The standard deviation and range of the absolute geometric distortion for the three studies were computed.

## 2.6 Measurement of geometric distortion

The phantom was scanned on three different MR-scanners at our institution using several different pulse sequences and acquisition parameters to demonstrate the use of the software and phantom to characterize geometric distortion. The demonstration of the phantom was divided into three evaluation studies. Imaging was performed in the axial scan plane for each of the studies. The vendor supplied 3D distortion correction was applied for all scans, except for a single scan, discussed below, for which the 2D distortion correction algorithm was used. Table 1 summarizes the acquisition parameters used for each of the studies.

Table 1: MR imaging acquisition parameters for three experimental studies.

| Experiment Number | 1 | 2 | 2 | 3 | 3 | 3 |
|---|---|---|---|---|---|---|
| MR Scanner Type | Diagnostic | Diagnostic | Diagnostic | Diagnostic | Diagnostic | MR-Linac |
| Protocol | T1 3D Axial | T1 3D Axial | T2 3D Axial | T2 3D Axial | T2 3D Axial | T2 3D Axial |
| Field strength (T) | 1.5 | 1.5 | 1.5 | 1.5 | 3.0 | 1.5 |
| TR (ms) | 8.5 | 8.5 | 1500.0 | 1500.0 | 2000.0 | 1535.0 |
| TE (ms) | 4.8 | 4.8 | 123.0 | 123.0 | 80.5 | 173.0 |
| Flip angle (°) | 12.0 | 12.0 | 160.0 | 160.0 | 90.0 | 90.0 |
| Bandwidth (Hz/voxel) | 200 | 200, 815 | 375, 750 | 375 | 391 | 394 |
| Number of averages | 1 | 1 | 2 | 2 | 1 | 2 |
| Acquisition matrix | 512 x 512 | 512 x 512 | 512 x 512 | 512 x 512 | 512 x 512 | 512 x 512 |
| Voxel size (mm³) | 1 x 1 x 2 | 1 x 1 x 2 | 1 x 1 x 2 | 1 x 1 x 2 | 1 x 1 x 2 | 1 x 1 x 2 |
| Distortion correction | 3D and 2D | 3D | 3D | 3D | 3D | 3D |

### 2.6.1 2D versus 3D distortion correction study

For the first experiment, T1-weighted images were acquired on a 1.5T diagnostic scanner with the 2D and 3D vendor supplied distortion correction applied separately. The purpose of the first validation study was to demonstrate the phantom correctly measures increased distortion for the 2D versus 3D distortion correction algorithm.

### 2.6.2 Bandwidth variation study

For the second experiment, the receiver bandwidth was varied from 200 to 815 Hz/pixel for T1-weighted imaging and 375 to 750 Hz/pixel for T2-weighted imaging. Geometric distortion is expected to decrease as the receiver bandwidth increases (Paulson *et al* 2015). All imaging was performed on the 1.5T diagnostic scanner for the bandwidth variation experiment. The objective of the second experiment was to demonstrate the software's ability to quantify geometric distortion for different acquisition parameters and to evaluate the sensitivity of the phantom to detect small changes in geometric distortion by varying the bandwidth.

*2.6.2 Geometric distortion versus MRI scanner type study*

The third experiment quantified geometric distortion for T1- and T2-weighted images acquired on three different scanners. The phantom was imaged on a 1.5T diagnostic scanner, a 1.5T MR-guided linear accelerator (Unity MR-Linac, Elekta AB, Stockholm, Sweden), and a 3.0T diagnostic scanner to demonstrate the phantom's compatibility with different scanner geometries. T2-weighted imaging parameters were selected to be similar (Table 1). However, it's important to note that the acquisition parameters were not selected to optimize image quality or minimize distortion across the three different scanners. For each of the described studies, the image volumes were analyzed using the developed software module.

## 3. Results

### 3.1 Evaluation of software accuracy

Table 2 presents the geometric distortion analysis performed using the vendor provided 3D Geometric QA Analysis tool (denoted *Ref.*) versus the software developed in this work (denoted *MDACC*). Differences in absolute distortion, specified as the 98th percentile of a histogram of distortion distances, between the vendor software results and the custom software results were 0.05 mm, 0.07 mm, 0.03 mm, and 0.01 mm for spherical volumes with diameters of 200 mm, 300 mm, 400 mm, and 500 mm defined about the isocenter. Similar differences may be observed for each of the ML, AP, and SI dimensions. For comparison, the differences were much smaller than the image pixel dimensions (1 mm) and the standard deviations of distortion errors (0.12 mm, 0.18 mm, 0.26 mm, and 0.55 mm) which indicates good agreement between the developed software and the reference results. The small differences may be attributed to several factors including how the software evaluates fiducials on the edge of the volume and potential differences in the implementation of the center-of-mass computation such as the size of the local region-of-interest used for the computations.

Table 2: Geometric distortion results for a reference vendor provided phantom and software (Ref.) versus the custom-software (MDACC) developed for the modular phantom.

| | \multicolumn{8}{c}{**Geometric Distortion (mm)**} | | | | | | | |
|---|---|---|---|---|---|---|---|---|
| | Ref. | MDACC | Ref. | MDACC | Ref. | MDACC | Ref. | MDACC |
| **Diameter** | 200 (mm) | | 300 (mm) | | 400 (mm) | | 500 (mm) | |
| **Absolute (98th %)** | 0.49 | 0.54 | 0.74 | 0.81 | 1.16 | 1.19 | 2.37 | 2.38 |
| **Absolute (Max)** | 0.50 | 0.77 | 0.85 | 1.01 | 1.60 | 1.46 | 7.90 | 5.88 |
| **ML (98th %)** | 0.32 | 0.30 | 0.54 | 0.41 | 0.80 | 0.62 | 1.60 | 0.93 |
| **ML (Max)** | 0.36 | 0.30 | 0.73 | 0.55 | 1.29 | 0.77 | 4.67 | 2.37 |
| **AP (98th %)** | 0.37 | 0.30 | 0.49 | 0.49 | 0.78 | 0.65 | 2.14 | 1.53 |
| **AP (Max)** | 0.39 | 0.37 | 0.54 | 0.63 | 1.23 | 1.08 | 7.30 | 2.94 |
| **SI (98th %)** | 0.34 | 0.41 | 0.58 | 0.58 | 0.81 | 0.76 | 1.23 | 0.79 |
| **SI (Max)** | 0.37 | 0.70 | 0.66 | 0.76 | 0.98 | 1.02 | 1.76 | 1.11 |

### 3.2 Reproducibility of phantom set-up

The standard deviation of the measured absolute geometric distortion across three independent phantom set-ups was 0.01 mm, 0.02 mm, and 0.07 mm for spherical regions-of-interest with diameters of 200 mm, 300 mm, and 400 mm, respectively. The range, defined as the difference between the maximum and minimum measured distortion, was 0.02 mm, 0.03 mm and 0.13 mm across the three imaging studies for the respective regions-of-interest. The small differences in geometric distortion observed for the different imaging studies can be attributed to a combination of set-up uncertainty and fiducial detection uncertainty. Thus, the portion of measured geometric distortion that can be attributed to variations in phantom set-up is expected to be on the order of 0.1 mm or less.

### 3.3 Measurement of geometric distortion

*3.3.1 2D versus 3D distortion correction study*

Geometric distortion was characterized for MR image volumes acquired with the 2D and 3D vendor supplied distortion correction algorithms applied (Table 1, experiment 1). Example heat maps for several 2D axial slices within the MR bore are shown in Figure 5. The x-axis (horizontal axis) of the heat maps corresponds to the medial-lateral dimension of the MR bore while the y-axis (vertical axis) corresponds to the anterior-posterior dimension. Each of the individual 2D heat maps corresponds to a 1-mm thick slice located at distances ranging from -100 mm to +100 mm, in the inferior-superior dimension, from isocenter. The solid line contains regions within the bore where the measured geometric distortion was less than or equal to 1.0 mm. The dashed line indicates regions with distortion less than or equal to 2.0 mm. The heat maps show an increase in the magnitude of geometric distortion as the distance from isocenter increases, as expected. Comparing the heat maps derived from 2D versus 3D distortion corrected images at the iso-plane (i.e. z = 0 mm) shows a similar distribution of measured geometric distortion. As the distance along the superior-inferior axis increases, however, an increase in geometric distortion is observed for the 2D distortion corrected images relative to the 3D distortion corrected images. As a result, the iso-lines corresponding to regions with geometric distortion less than 1.0 mm (solid line) and 2.0 mm (dashed line) encompass a greater region for the 3D distortion corrected images.

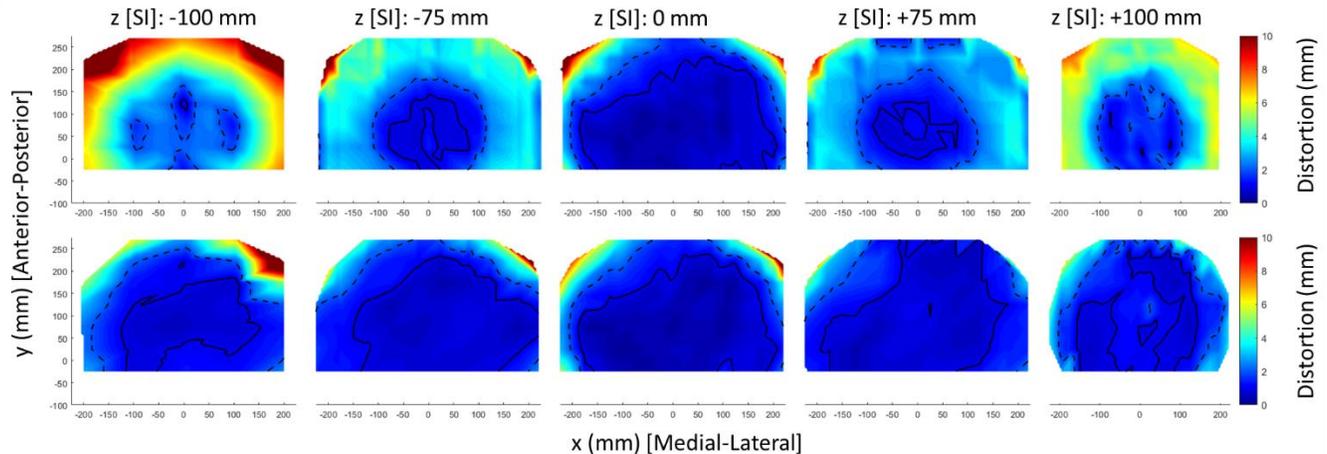

Figure 5: Heat maps present the absolute geometric distortion measured in 2D axial planes ranging from -100 mm inferior of isocenter (left) to +100 mm superior of isocenter with 2D (top row) and 3D (bottom row) vendor supplied distortion correction applied. The solid line indicates regions in which distortion was less than or equal to 1.0 mm. The dashed line indicates regions in which distortion was less than or equal to 2.0 mm.

Table 3 presents the mean, standard deviation, and maximum distortion values for the 2D versus 3D vendor supplied distortion correction study. Results are presented in terms of the absolute geometric distortion as well as the component of the absolute distortion occurring in each of the ML, AP, and SI dimensions. Figure 6 presents a series of scatter plots showing the distributions of measured geometric distortion for each of the phantom's reference points. Both Table 3 and Figure 6 demonstrate a reduction in geometric distortion when the 3D distortion correction algorithm is applied relative to the 2D algorithm, as expected. Figure 6D and Table 3 show the increase in absolute geometric distortion can primarily be attributed to increased distortion along the superior-inferior direction with the 2D versus 3D correction algorithm.

Table 3: Geometric distortion with 3D and 2D vendor supplied distortion correction applied for spherical volumes-of-interest defined about magnetic isocenter. Distortion is specified as, $\mu \pm \sigma$ (x), where $\mu$ indicates the mean, $\sigma$ the standard deviation, and x the maximum value of geometric distortion for a distribution of measurement points.

|  | Diameter of spherical volume-of-interest (mm) | | | |
|---|---|---|---|---|
|  | 200 | 300 | 400 | 500 |
| **Absolute** | | | | |
| **3D Correction** | 0.53 ± 0.23 (1.06) | 0.67 ± 0.30 (1.51) | 0.74 ± 0.36 (2.87) | 1.04 ± 0.74 (6.22) |
| **2D Correction** | 0.65 ± 0.35 (1.61) | 1.18 ± 0.89 (4.24) | 1.50 ± 1.12 (6.04) | 2.09 ± 1.71 (8.05) |
| Medial-Lateral | | | | |
| 3D Correction | 0.01 ± 0.25 (0.54) | 0.06 ± 0.37 (0.91) | 0.00 ± 0.48 (2.85) | -0.07 ± 0.88 (3.91) |
| 2D Correction | 0.00 ± 0.25 (0.50) | 0.02 ± 0.39 (0.98) | -0.02 ± 0.50 (2.93) | -0.08 ± 0.87 (3.88) |
| Anterior-Posterior | | | | |
| 3D Correction | -0.06 ± 0.43 (0.79) | -0.07 ± 0.54 (1.25) | -0.14 ± 0.57 (1.28) | 4.80 ± 0.79 (4.80) |
| 2D Correction | -0.05 ± 0.43 (0.79) | -0.04 ± 0.57 (1.78) | -0.12 ± 0.57 (1.78) | 4.83 ± 0.77 (4.83) |
| Superior-Inferior | | | | |
| 3D Correction | 0.17 ± 0.24 (0.80) | 0.05 ± 0.31 (1.20) | 0.04 ± 0.32 (1.20) | -0.03 ± 0.46 (1.78) |
| 2D Correction | 0.18 ± 0.53 (1.47) | 0.07 ± 1.31 (3.83) | 0.15 ± 1.70 (5.96) | 0.22 ± 2.42 (7.45) |

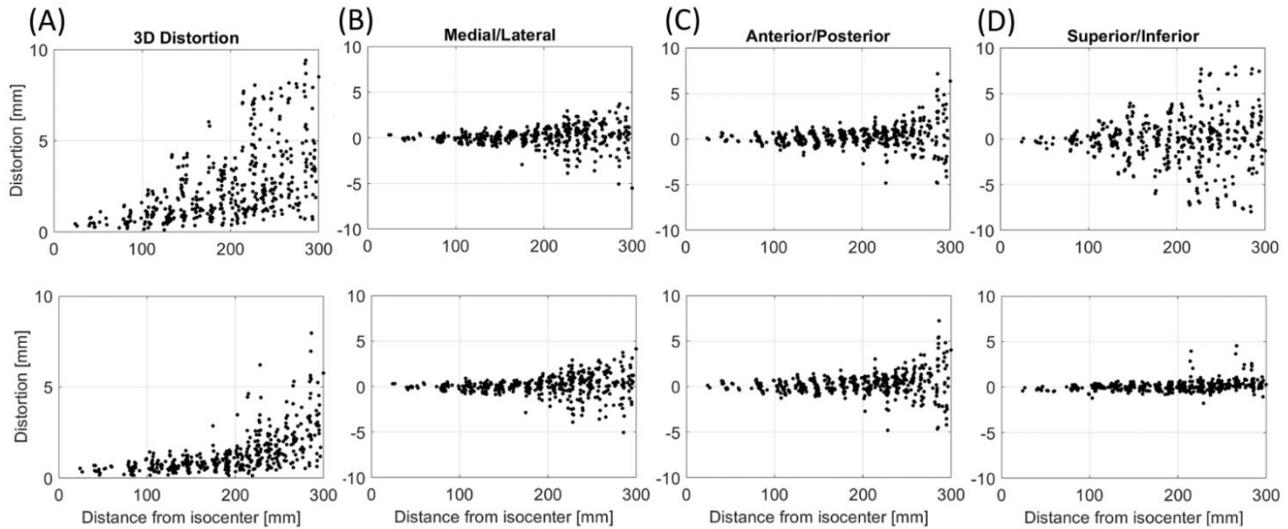

Figure 6: Geometric distortion measured for a collection of points is shown with 2D (top row) and 3D (bottom row) vendor supplied geometric distortion correction applied. Distortion is specified in terms of the absolute distortion (column A), and components in the medial-lateral (column B), anterior-posterior (column C) and superior-inferior (column D) dimensions.

*3.3.2 Bandwidth variation study*

The second validation study quantified geometric distortion for T1- and T2-weighted imaging with the receiver bandwidth varied from 200 to 815 Hz/voxel and 375 to 750 Hz/voxel, respectively. Statistics for the bandwidth variation study are presented in Table 4. The mean geometric distortion was observed to decrease slightly from 1.5 mm to 1.4 mm as the receiver bandwidth increased from 200 Hz/voxel to 815 Hx/voxel. Similarly, the maximum distortion decreased from 5.6 mm to 5.1 mm. Mean distortion measured 1.5 mm at both 375 Hz/voxel and 750 Hz/voxel for T2-weighted imaging. The maximum distortion decreased from 5.9 mm to 5.7 mm. Geometric distortion was successfully measured using the developed software for each combination of pulse sequence and receiver bandwidth.

Table 4: Geometric distortion versus receiver bandwidth for T1-weighted and T2-weighted pulse sequences. Distortion is specified as, μ ± σ (x), where μ indicates the mean, σ the standard deviation, and x the maximum value of geometric distortion for a distribution of measurement points.

| Diameter (mm) [inner-outer] | T1-Weighted | | T2-Weighted | |
|---|---|---|---|---|
| | 200 Hz/voxel | 815 Hz/voxel | 375 Hz/voxel | 750 Hz/voxel |
| 0-200 | 0.5 ± 0.3 (1.1) | 0.5 ± 0.2 (1.0) | 0.5 ± 0.2 (0.9) | 0.5 ± 0.2 (0.8) |
| 200-300 | 0.7 ± 0.4 (2.1) | 0.7 ± 0.4 (1.9) | 0.8 ± 0.4 (1.9) | 0.7 ± 0.3 (1.9) |
| 300-400 | 1.2 ± 0.6 (3.1) | 1.1 ± 0.6 (3.5) | 1.2 ± 0.6 (3.5) | 1.2 ± 0.5 (3.0) |
| >400 | 2.5 ± 1.7 (5.6) | 2.2 ± 1.5 (5.1) | 2.5 ± 2.6 (5.9) | 2.5 ± 2.8 (5.7) |
| *>0* | *1.5 ± 1.4* | *1.4 ± 1.2* | *1.5 ± 1.8* | *1.5 ± 1.9* |

### 3.3.3 Geometric distortion versus MRI scanner type study

The third validation study compared geometric distortion for T2-weighted images of the phantom acquired on three different scanners. Figure 7 presents three-dimensional heat maps for each of the scanners studied. Geometric distortion increased as the distance from iso-center increased, as expected. The MR-Linac has a split-coil design (Raaymakers *et al* 2009). However, this modification has been achieved with minimal or no degradation to the spatial fidelity. The mean geometric distortion as well as the standard deviation and maximum values for the distribution of measurement points are summarized in Table 5. The absolute geometric distortion is broken down into the component of distortion along the ML, AP, and SI dimensions for each of the scanners and displayed in Figure 8 as a series of histogram plots. The MR-Linac provided images with geometric distortion less than or comparable to both diagnostic scanners. For a 400 mm diameter spherical volume-of-interest the mean geometric distortion was 0.94 ± 0.58 mm for the MR-Linac versus 0.90 ± 0.53 mm for the 1.5 T scanner and 1.15 ± 0.62 mm for the 3.0T scanner. The mean geometric distortion was slightly greater for each volume-of-interest analyzed for the 3.0T scanner compared to the two 1.5T scanners.

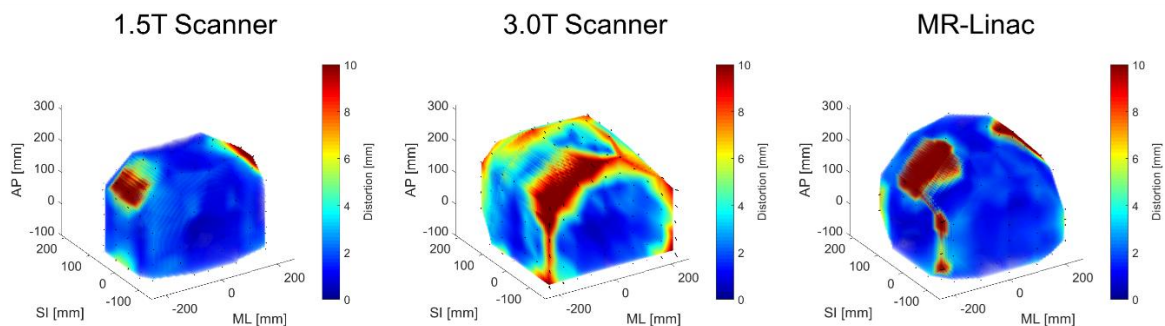

Figure 7: Three-dimensional heat maps present geometric distortion measured within three different MR bores. Geometric distortion was measured for a diagnostic 1.5 T scanner (left), a diagnostic 3.0 T scanner (center), and a 1.5T MR-Linac (right). Acquisition parameters for each scanner are summarized in Table 1.

Table 5: Geometric distortion for three different MRI scanners. Distortion is specified as, μ ± σ (x), where μ indicates the mean, σ the standard deviation, and x the maximum value of geometric distortion for a distribution of measurement points.

| MRI Scanner | Diameter of spherical volume-of-interest (mm) | | | |
|---|---|---|---|---|
| | 200 | 300 | 400 | 500 |
| 1.5T Scanner | 0.51 ± 0.19 (0.93) | 0.69 ± 0.34 (1.87) | 0.90 ± 0.53 (3.47) | 1.22 ± 0.86 (5.86) |
| 3.0T Scanner | 0.75 ± 0.38 (1.73) | 0.87 ± 0.39 (2.33) | 1.15 ± 0.62 (3.43) | 1.62 ± 1.05 (6.97) |
| MR-Linac | 0.52 ± 0.32 (1.33) | 0.73 ± 0.43 (1.94) | 0.94 ± 0.58 (3.78) | 1.09 ± 0.69 (4.63) |

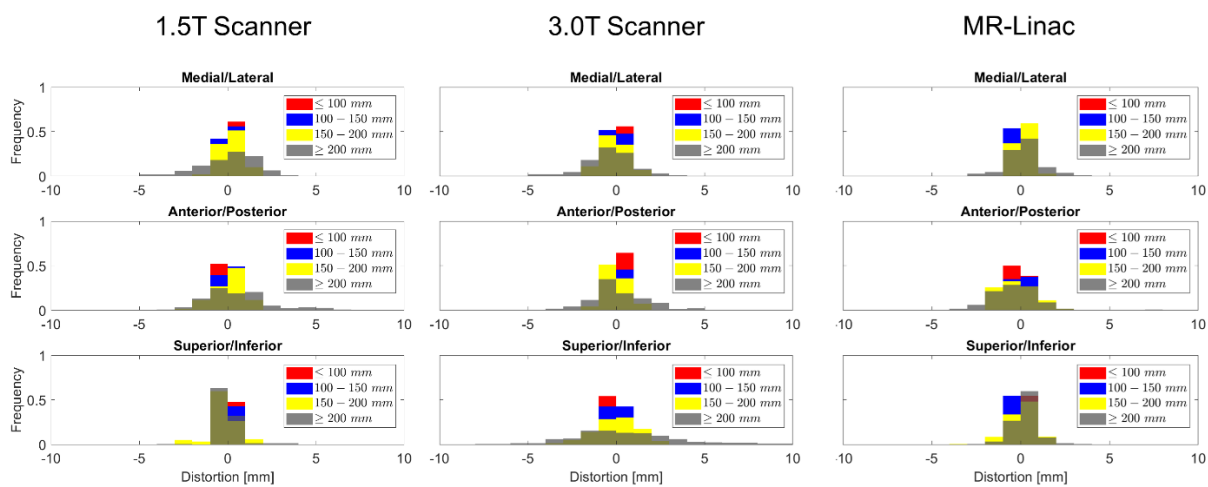

Figure 8: Histograms of geometric distortion within the medial-lateral (top row), anterior-posterior (middle row), and superior-inferior (bottom row) dimensions are shown for three different MRI scanners, including an MR-Linac.

## 4. Discussion

A modular *block* phantom and software application were developed to quantify geometric distortion in MR-guided radiation therapy and simulation. Several validation studies performed on three different MRI scanners, including the Unity MR-Linac, demonstrated the capabilities of the phantom. In the first study, MR images were acquired with the 2D and 3D vendor supplied distortion correction algorithms applied separately. As expected, geometric distortion was reduced with the 3D correction algorithm applied when compared to the 2D correction algorithm. As also expected, the measurements demonstrated the 3D correction algorithm effectively reduced distortion along the superior-inferior axis. The maximum distortion along the superior-inferior axis decreased from 7.45 mm to 1.78 mm for a 500 mm diameter spherical volume-of-interest, with the 2D and 3D correction algorithms applied, respectively. Thus, the first study demonstrated the phantom can be used to effectively measure the individual components of geometric distortion along each of the cardinal axes. The receiver bandwidth was varied in the second experiment. The measured results showed a decrease in the maximum geometric distortion from 5.6 mm to 5.1 mm when the bandwidth was increased from 200 Hz/voxel to 815 Hz/voxel for T1-weighted imaging, and from 5.9 mm to 5.7 mm as the bandwidth was increased from 375 Hz/voxel to 750 Hz/voxel for T2-weighted imaging. Finally, a third experiment was performed to demonstrate the phantom can be used with different scanner types. The phantom was scanned on three different scanners and successfully used to quantify geometric distortion in each of the different bores.

One limitation of the phantom design is that care must be taken to ensure a reproducible set-up. Deviations in phantom set-up from the reference model would produce errors in the distortion measurements. The phantom was designed with markings and labels to align each of the modular blocks and ensure a reproducible set-up. The standard deviation of geometric distortion across all imaging studies performed with independent phantom set-ups was less than 0.1 mm. The small degree of uncertainty demonstrates that the phantom set-up is reproducible. Nevertheless, a modified version of the phantom, currently under construction, will include positioning rods that bind the blocks together, similar to the work by *Price et al* (2017). Nevertheless, integrated non-modular commercial phantoms may be preferred for routine QA by clinical physicists that do not need the added flexibility and modular design of the phantom required by researchers.

For each of the imaging studies performed, no efforts were made to find acquisition parameters that minimized geometric distortion. Increasing the receiver bandwidth reduces the image acquisition time but is also known to decrease the signal-to-noise ratio (SNR) which may reduce lesion visibility. The successful development and validation of the described modular phantom will enable future studies to understand spatial distortions at a given bandwidth and weigh the trade-offs in loss of SNR versus distortion when designing new protocols. The development of the phantom will also will enable future planned work to survey geometric distortion across all of the MRI scanners at our institution. Future efforts to improve the phantom may concentrate on improving the segmentation algorithm to remove the need for manual review. One approach may be to train a convolutional neural network using reference fiducial positions determined with the software described here.

## 5. Conclusion

A novel phantom and software to characterize geometric distortion within MR images was developed. The phantom was designed to span a large field-of-view for quality assurance of scanners used for MRI simulation and MR-guided radiotherapy. The modular phantom design makes the phantom portable and compatible with a wide variety of bore geometries and RF coils.

## Disclosures

Dr. Fuller and Dr. Wang received funding and salary support related to this project from the National Institutes of Health (NIH), including: the National Institute for Dental and Craniofacial Research Establishing Outcome Measures Award (1R01DE025248/R56DE025248) and an Academic Industrial Partnership Grant (R01DE028290); NCI Early Phase Clinical Trials in Imaging and Image-Guided Interventions Program  (1R01CA218148); an NIH/NCI Cancer Center Support Grant (CCSG) Pilot Research Program Award from the UT MD Anderson CCSG Radiation Oncology and Cancer Imaging Program (P30CA016672) and an NIH/NCI Head and Neck Specialized Programs of Research Excellence (SPORE) Developmental Research Program Award (P50 CA097007). Dr. Fuller received funding and salary support unrelated to this project from: National Science Foundation (NSF), Division of Mathematical Sciences, Joint NIH/NSF Initiative on Quantitative Approaches to Biomedical Big Data (QuBBD) Grant (NSF 1557679); NSF Division of Civil, Mechanical, and Manufacturing Innovation (CMMI) standard grant (NSF 1933369) a National Institute of Biomedical Imaging and Bioengineering (NIBIB) Research Education Programs for Residents and Clinical Fellows Grant (R25EB025787-01); the NIH Big Data to Knowledge (BD2K) Program of the National Cancer Institute (NCI) Early Stage Development of Technologies in Biomedical Computing, Informatics, and Big Data Science Award (1R01CA214825). Direct infrastructure support was provided by the multidisciplinary Stiefel Oropharyngeal Research Fund of the University of Texas MD Anderson Cancer Center Charles and Daneen Stiefel Center for Head and Neck Cancer and the Cancer Center Support Grant (P30CA016672) and the MD Anderson Program in Image-guided Cancer Therapy. Dr. Fuller has received direct industry grant support, honoraria, and travel funding from Elekta AB.